\documentclass[a4paper,fleqn]{cas-dc}
\usepackage{graphicx,amssymb,amsmath}
\usepackage{setspace}
\usepackage[authoryear]{natbib}
\usepackage{xcolor}
\usepackage{soul}
\sethlcolor{yellow}
\AtBeginDocument{} 

\begin{document}

\shorttitle{Extinction for Nearly 90 Million Star}
\shortauthors{Sun et al.}

\title[mode=title]{Interstellar Extinction Estimates for Nearly 90 Million Stars Based on Synthetic Photometry from Gaia XP Spectra}

\author[1]{Bao-Kun Sun}
\cormark[1]
\ead{sunbaokun@cxtc.edu.cn}

\author[1]{Yan-Hui Chen}

\author[1]{Zheng Yu}
\cormark[1]
\ead{yuzheng_hs@163.com}

\author[2]{Bai-Song Zhang}
\cormark[1]
\ead{zhangbaisong@mail.ynu.edu.cn}

\cortext[1]{Corresponding author.}

\affiliation[1]{organization={Institute of Astrophysics, Chuxiong Normal College}, city={Chuxiong}, postcode={675000}, country={China}}

\affiliation[2]{organization={School of Integrated Circuits, Dezhou Vocational and Technical College}, city={Dezhou}, postcode={253034}, country={China}}

\begin{abstract}
Accurate stellar extinction measurements are essential for recovering intrinsic stellar properties and investigating the three-dimensional distribution of interstellar dust. We present reddening estimates for 87,270,425 stars using synthetic SMSS $gri$ photometry derived from corrected Gaia DR3 XP spectra, combined with Gaia BP and RP, 2MASS $JHK_s$, and WISE $W_1$ photometry. Stellar extinction is determined through empirical SED fitting based on reference stellar loci constructed separately for dwarfs and giants. Comparisons with three external extinction catalogs show good overall agreement, with dispersions of $0.049$--$0.075\,{\rm mag}$ in $E(B-V)$. The catalog covers approximately $99.99\%$ of the sky. Using stars with $\sigma_d/d<0.2$, we further construct reddening maps in four distance ranges from 0 to 5\,kpc, revealing prominent molecular clouds and large-scale Galactic dust structures. Extinction-corrected color--absolute-magnitude diagrams show significantly tighter stellar sequences and consistent red-clump features in different Galactic fields. The resulting catalog provides a homogeneous data set for stellar extinction correction and studies of Galactic dust structure.
\end{abstract}



\maketitle

\section{Introduction}
\label{sec:introduction}

Interstellar dust absorbs and scatters stellar radiation, altering the observed fluxes and colors of stars and producing interstellar extinction and reddening \citep{draine2003interstellar}. Accurate correction for these effects is essential for recovering the intrinsic properties of stars and for deriving reliable stellar parameters \citep[e.g.,][]{sun2025photometric, green2021data, zhang2023parameters, sun2023spar}. Large samples of stars with well-determined extinction are also of fundamental importance for studies of Galactic stellar populations, structure, and evolution \citep[e.g.,][]{chen2017constraining, chen2019three, an2024blueprint, sun2025stellar}.

Traditionally, extinction information has been provided by two-dimensional (2D) dust maps, such as the widely used Schlegel, Finkbeiner, and Davis (SFD) map \citep{schlegel1998maps} and the Planck dust maps \citep{abergel2013planck}. These maps measure the integrated extinction along a given line of sight but contain no distance information, which limits their application to stars located within the Galactic dust layer. With the advent of large-scale photometric, spectroscopic, and astrometric surveys, extinction can now be estimated for individual stars, enabling the three-dimensional distribution of interstellar dust to be investigated. A number of large stellar extinction catalogs and three-dimensional dust maps have consequently been constructed using different combinations of optical and infrared photometry, spectroscopy, and Gaia astrometry \citep{green20193d, chen2019three, guo2021three}. In particular, multi-band spectral energy distribution (SED) fitting provides an efficient approach for determining extinction for large stellar samples \citep{berry2012milky, chen2014three, guo2021three}.

The release of Gaia DR3 has opened a new avenue for large-scale photometric studies. In addition to its astrometric and integrated photometric measurements, Gaia DR3 provides flux-calibrated low-resolution BP/RP (XP) spectra for approximately 220 million sources \citep{vallenari2023gaia,de2023gaia}. An important advantage of these spectra is that synthetic photometry can be generated for suitable photometric passbands contained within the XP wavelength range \citep{gaia2023gaia}. This capability effectively transforms the XP spectral database into a flexible, nearly all-sky photometric survey in a variety of existing or customized photometric systems. The broad sky coverage and enormous sample size of Gaia XP therefore provide an opportunity to extend photometric techniques far beyond the original footprints of ground-based imaging surveys.

The scientific potential of Gaia XP spectra and their synthetic-photometry products has been demonstrated in a variety of applications. \citet{de2023gaia} describe the processing and validation of the Gaia DR3 XP spectra, while \citet{gaia2023gaia} present the methodology and applications of synthetic photometry derived from these spectra. Subsequent studies have applied Gaia XP information to photometric recalibration \citep{chen2024photometric}, nearly all-sky identification of blue horizontal-branch stars using synthetic SkyMapper Southern Survey (SMSS; \citealt{wolf2018skymapper}) colors \citep{hu2025identification}, and empirical calibration of red-clump absolute magnitudes in the Multi-channel Photometric Survey Telescope (Mephisto; \citealt{chen2024early}) and the Chinese space station survey telescope(CSST; \citealt{csst2026introduction}) photometric systems \citep{yu2025empirically}. These applications highlight the broad utility of Gaia XP spectra for large-scale photometric and stellar studies. Meanwhile, systematic effects in the Gaia XP spectra associated with source color, magnitude, reddening, and wavelength have been recalibrated by \citet{huang2024comprehensive}, further improving their suitability for precision synthetic photometry.

Motivated by these developments, in this work we use the corrected Gaia XP spectra to construct synthetic SMSS $g$, $r$, and $i$ photometry and combine these measurements with optical and infrared photometry from Gaia, the Two Micron All Sky Survey (2MASS; \citealt{skrutskie2006two}), and the Wide-field Infrared Survey Explorer (WISE; \citealt{wright2010wide}). We then apply an empirical SED-fitting method to determine the extinction of individual stars. A reference SED library is constructed from a large sample of low-extinction stars, and the observed multi-band colors of each target are fitted against the corresponding intrinsic stellar locus. After applying stringent photometric quality criteria, we obtain extinction estimates for approximately 87 million stars. The resulting catalog provides a large and homogeneous set of stellar extinction measurements based on Gaia XP synthetic photometry and offers a valuable resource for extinction correction, Galactic structure studies, and future investigations of the three-dimensional distribution of interstellar dust.

The remainder of this paper is organized as follows. Section~\ref{sec:data_methods} describes the input data, sample selection, extinction-estimation method, and resulting catalog and maps. Section~\ref{sec:validation and application} presents comparisons with previous studies and an application to extinction-corrected color--absolute-magnitude diagrams. Section~\ref{sec:conclusions} summarizes our main results.

\section{Data and Methods}
\label{sec:data_methods}

\subsection{Photometric Data and Sample Selection} \label{subsec:photometric data and sample selection}

Gaia DR3 \citep{vallenari2023gaia} provides low-resolution BP/RP (XP) spectra for approximately 220 million sources, spanning a wavelength range of 336--1020 nm. Systematic errors in the XP spectra associated with source color, magnitude, reddening, and wavelength were recently recalibrated by \citet{huang2024comprehensive} using well-calibrated spectrophotometric data from CALSPEC \citep{bohlin2014techniques}, the Hubble Next Generation Spectral Library (NGSL; \citealt{heap2007hubble}), and the Large Sky Area Multi-Object Fiber Spectroscopy Telescope (LAMOST; \citealt{zhao2012lamost, liu2013lss}). These corrections significantly improve the spectrophotometric accuracy of the XP spectra, particularly in the near-ultraviolet and around the BP/RP junction. Throughout this work, we adopt the corrected XP spectra from \citet{huang2024comprehensive} to construct synthetic photometry for stellar extinction estimation.

To derive synthetic photometry from the corrected XP spectra, we integrate each spectrum over the transmission curves of SMSS photometric system. The transmission curves were obtained from the Spanish Virtual Observatory (SVO) Filter Profile Service \citep{rodrigo2012svo,rodrigo2020svo}. We adopt the SMSS $g$, $r$, and $i$ passbands to construct the synthetic magnitudes. These broad optical passbands are well covered by the wavelength range of the Gaia XP spectra and provide complementary optical wavelength coverage to the Gaia BP and RP bands.

\begin{figure*}[t]
\centering
\includegraphics[width=0.95\textwidth]{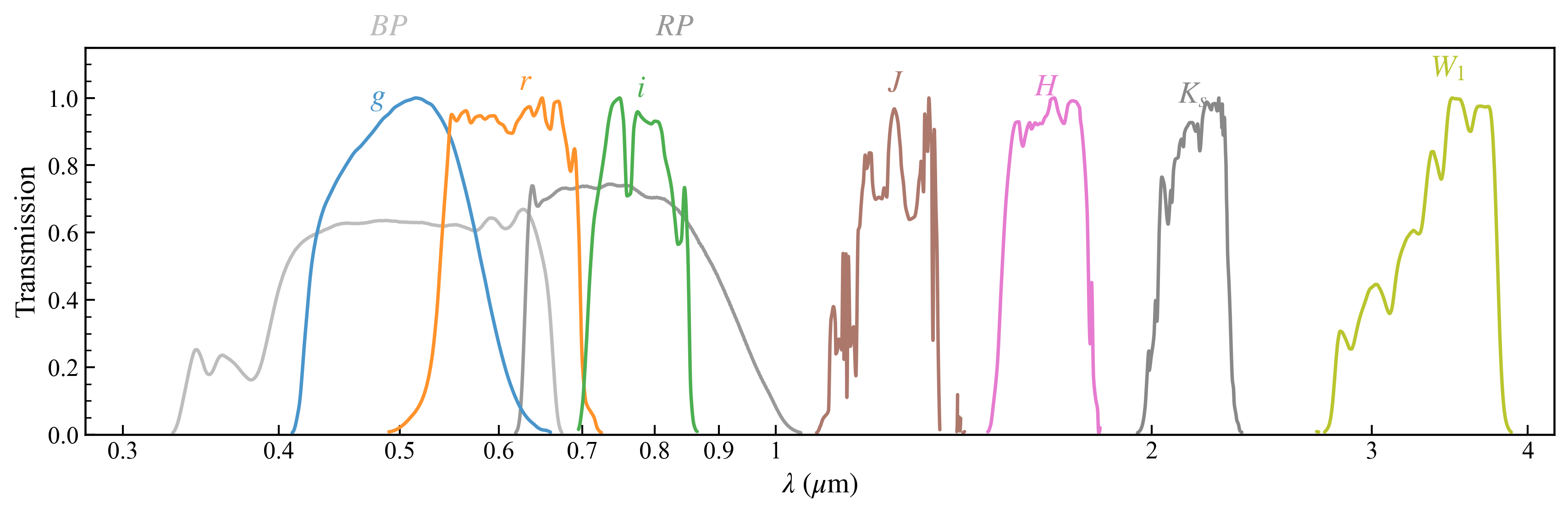}
\caption{Transmission curves of the photometric passbands adopted in this work, including the Gaia BP and RP bands, the SMSS $g$, $r$, and $i$ bands, the 2MASS $J$, $H$, and $K_s$ bands, and the WISE $W_1$ band. The Gaia transmission curves are taken from \citet{riello2021gaia}, the SMSS transmission curves from \citet{wolf2018skymapper}, the 2MASS transmission curves from \citet{cohen2003spectral}, and the WISE transmission curve from \citet{wright2010wide}. The transmission curves were obtained through the SVO Filter Profile Service \citep{rodrigo2012svo,rodrigo2020svo}.}
\label{fig:transmission_curves}
\end{figure*}

In addition to the synthetic $g$, $r$, and $i$ photometry derived from the corrected Gaia XP spectra, we also adopt Gaia DR3 BP and RP photometry, together with near-infrared photometry from 2MASS and mid-infrared photometry from WISE. Gaia provides homogeneous, high-precision broad-band photometry for a very large number of sources across the sky \citep{vallenari2023gaia, hambly2021gaia}. 2MASS provides all-sky photometry in the $J$, $H$, and $K_s$ bands, reaching approximately 10-$\sigma$ point-source depths of 15.8, 15.1, and 14.3 mag (Vega), respectively, with an overall photometric calibration accuracy of about 0.02 mag \citep{skrutskie2006two}. WISE surveyed the entire sky in four mid-infrared bands centered at approximately 3.4, 4.6, 12, and 22~$\mu$m ($W_1$, $W_2$, $W_3$, and $W_4$, respectively), of which we adopt the $W_1$-band photometry from the AllWISE Source Catalog.
The $W_1$ photometry reaches an approximate 5-$\sigma$ point-source depth of 16.9 mag (Vega) in unconfused regions \citep{cutri2013explanatory}.
The absolute photometric calibration uncertainty in this band is approximately 0.03 mag.
The $W_2$, $W_3$, and $W_4$ bands are not used owing to photometric quality limitations in our sample. The resulting photometric data set therefore consists of nine bands: $g$, $r$, $i$, BP, RP, $J$, $H$, $K_s$, and $W_1$. The transmission curves of these passbands are shown in Fig.~\ref{fig:transmission_curves}. Together, these bands provide broad wavelength coverage from the optical to the mid-infrared, offering strong constraints on stellar extinction.

To ensure the quality of the photometric sample, we require photometric uncertainties smaller than 0.02 mag in the $g$, $r$, $i$, BP, and RP bands, and smaller than 0.08 mag in the $J$, $H$, $K_s$, and $W_1$ bands. After applying these quality cuts, our final sample contains 87,270,425 stars.

\subsection{Extinction Estimation}
\label{sec:extinction}

To estimate the extinction of individual stars, we adopt a SED fitting method similar to those of \citet{berry2012milky}, \citet{chen2014three} and \citet{guo2021three}. In this approach, we assume that the intrinsic optical and infrared colors of a star are primarily determined by its effective temperature, while the observed colors are additionally affected by interstellar extinction. The effects of surface gravity and metallicity on stellar colors are neglected. Therefore, an empirical reference SED library based on low-reddening reference stars is required for the subsequent extinction fitting. To construct this library, we select stars with photometric uncertainties smaller than 0.01\,mag in the $g$, $r$, $i$, BP, and RP bands and smaller than 0.05 mag in the $J$, $H$, $K_s$, and $W_1$ bands. We further require $0.86E(B-V)_{\rm SFD}<0.02$ mag to select stars in low-extinction regions and minimize reddening-induced shifts and scatter in the stellar loci used to construct the reference SED library.
The factor of 0.86 applies a 14\% downward recalibration to the original SFD reddening scale, as supported by comparisons with stellar colors \citep{schlafly2010blue, schlafly2011measuring, yuan2013empirical}.
We also adopt stars with relative distance uncertainties of $\sigma_d/d<0.2$ to limit the uncertainties in the absolute magnitudes used to distinguish giants from dwarfs, thereby reducing misclassification between the two populations.
After applying these criteria, a total of 1,755,077 stars are retained for constructing the reference SED library.

\begin{figure*}[t]
\centering
\includegraphics[width=\textwidth]{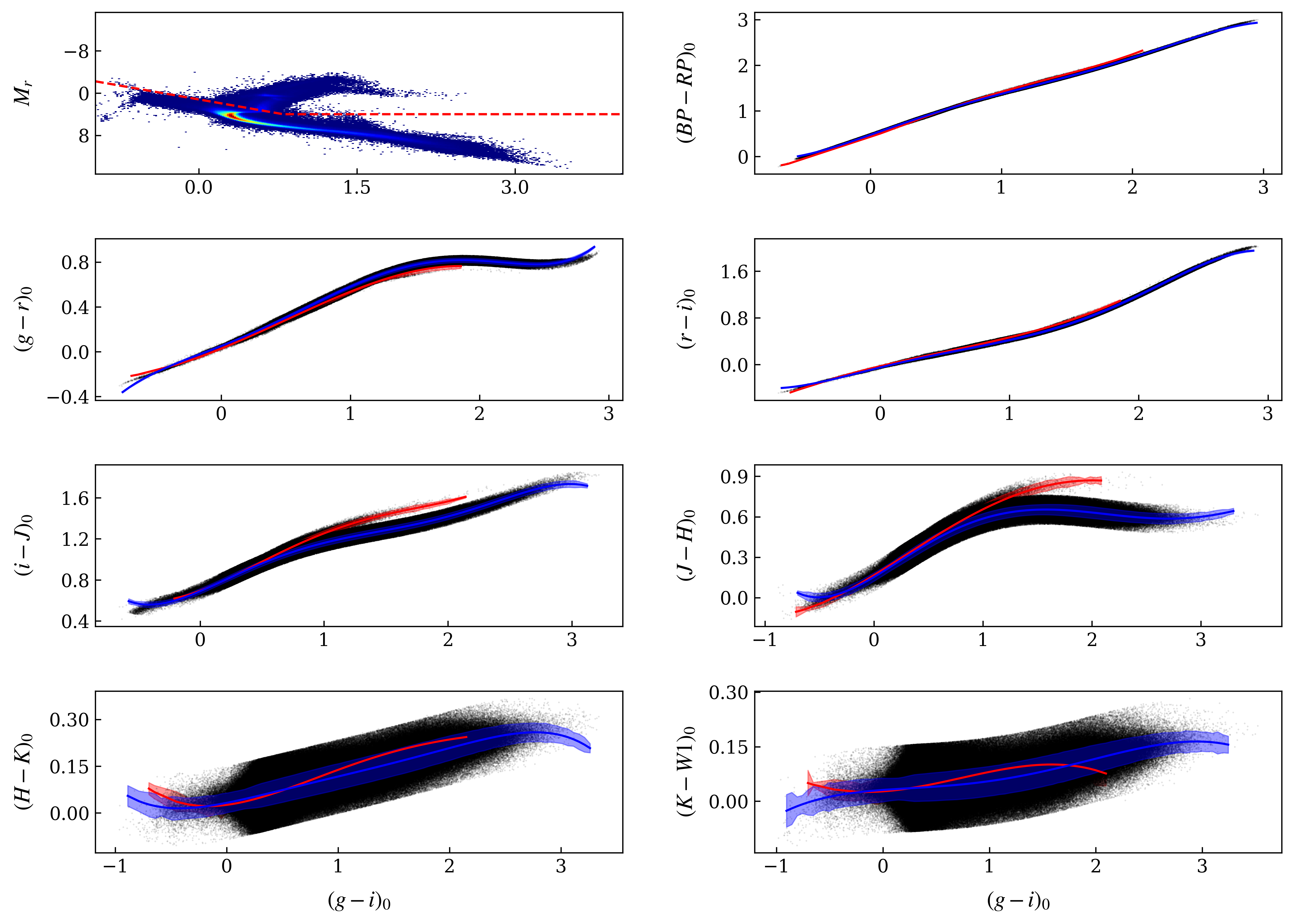}
\caption{Fits of the reference stellar loci using the synthetic SMSS $g$, $r$, and $i$ photometry derived from Gaia XP spectra, together with Gaia, 2MASS, and WISE photometry. Approximately 1.7 million low-extinction stars are used to construct the reference stellar loci. The top-left panel shows the color--absolute-magnitude diagram, with the absolute $r$-band magnitude $M_r$ plotted against the intrinsic color $(g-i)_0$. The red dashed line marks the adopted boundary separating giants above the line (smaller $M_r$) from dwarfs below it. The remaining panels show the intrinsic color--color distributions as a function of $(g-i)_0$, with the fitted stellar loci for giants and dwarfs overplotted as red and blue solid curves, respectively.}
\label{fig:stellar_locus}
\end{figure*}

The photometry of the selected reference stars is corrected for interstellar extinction before the reference stellar loci are constructed. Since dwarfs and giants exhibit different color--color relations, we divide the reference sample into dwarf and giant subsamples based on their positions in the color--absolute-magnitude diagram \citep{guo2021three,sun2025photometric}, as illustrated in the top-left panel of Fig.~\ref{fig:stellar_locus}, and fit their stellar loci separately. This diagram shows the absolute $r$-band magnitude $M_r$ as a function of the intrinsic color $(g-i)_0$. The adopted separation boundary is indicated by the red dashed line: stars above the boundary are classified as giants, while those below it are classified as dwarfs. The remaining panels of Fig.~\ref{fig:stellar_locus} show the intrinsic color--color relations parameterized by $(g-i)_0$ for the dwarf and giant samples, together with their corresponding best-fitting stellar loci. We adopt fifth-order and fourth-order polynomials to fit the stellar loci of dwarfs and giants, respectively. During the fitting procedure, two successive iterations of $3\sigma$ clipping are performed to remove outliers and obtain robust stellar loci. The best-fitting polynomial coefficients are listed in Table~\ref{tab:locus}. The resulting relations define the intrinsic colors adopted in the reference SED library. The typical fitting residuals for the optical colors are approximately 0.011--0.013\,mag for both dwarfs and giants, while those for the near-infrared colors are approximately 0.02--0.04\,mag.
The somewhat larger residuals for dwarfs, particularly in colors involving infrared bands, may partly reflect differences in the apparent-magnitude and signal-to-noise distributions of the two subsamples, since infrared photometric uncertainties increase toward fainter magnitudes.
These residuals include contributions from both photometric uncertainties and intrinsic stellar color variations, and characterize the scatter around the fitted mean stellar loci.

For each star, the extinction is then determined by comparing its observed multi-band colors with the intrinsic colors predicted by the corresponding reference stellar locus. For a color formed by two bands, $\lambda_1$ and $\lambda_2$, the reddened model color can be expressed as

\begin{equation}
c_{\rm mod}
=
c_{\rm sl}(C_0)
+
(C_{\lambda_1}-C_{\lambda_2})A_r,
\label{eq:sedcolor}
\end{equation}

where $c_{\rm mod}$ is the model color, $c_{\rm sl}(C_0)$ is the intrinsic color predicted by the reference stellar locus at a given intrinsic reference color $C_0$, $A_r$ is the extinction in the $r$ band, and $C_{\lambda}=A_{\lambda}/A_r$ is the relative extinction coefficient for band $\lambda$. For the extinction law, we adopt the extinction coefficients from \citet{huang2019milky} and \citet{chen2019three}, namely $R_{\lambda}=A_{\lambda}/E(B-V)=3.407$, 2.685, 2.030, 0.820, 0.520, 0.350, 0.210, 3.240, and 1.910 for the $g$, $r$, $i$, $J$, $H$, $K_s$, $W_1$, BP, and RP bands, respectively. The relative extinction coefficients used in Equation~\ref{eq:sedcolor} are therefore calculated as $C_{\lambda}=R_{\lambda}/R_r$, with $R_r=2.685$. We simultaneously fit all adopted colors by minimizing the reduced chi-square statistic

\begin{equation}
\chi_{\nu}^{2} =
\frac{1}{N_{\rm col}-2}
\sum_{i=1}^{N_{\rm col}}
\left(
\frac{c_{{\rm obs},i}-c_{{\rm mod},i}}
{\sigma_i}
\right)^2 ,
\label{eq:chi2}
\end{equation}

where $c_{{\rm obs},i}$ and $c_{{\rm mod},i}$ are the observed and model colors, respectively, and $\sigma_i$ denotes the corresponding uncertainty, including contributions from both the photometric errors and the fitting uncertainty of the reference stellar locus. We use $N_{\rm col}=7$ colors and fit two parameters, the intrinsic reference color $C_0$ and $A_r$, giving $\nu=N_{\rm col}-2=5$ degrees of freedom. The values corresponding to the minimum $\chi_{\nu}^{2}$ are adopted as the best-fitting parameters for each star. To estimate the extinction uncertainty, we use the unnormalized statistic $\chi^2=\nu\chi_{\nu}^{2}$. For each trial value of $A_r$, we minimize $\chi^2$ over $C_0$ to obtain the one-dimensional profile $\chi^2_{\rm prof}(A_r)$. The $1\sigma$ confidence interval of $A_r$ is defined by $\Delta\chi^2=1$ relative to the minimum of this profile, allowing for asymmetric lower and upper uncertainties. We adopt the mean of the two interval widths as the uncertainty in $A_r$ and convert it to the reddening uncertainty according to

\begin{equation}
\sigma_{E(B-V)}
=
\frac{\sigma_{A_r,-}+\sigma_{A_r,+}}
{2R_r},
\label{eq:ebv_error}
\end{equation}

where $\sigma_{A_r,-}$ and $\sigma_{A_r,+}$ are the lower and upper
uncertainties in $A_r$, respectively.

\begin{table}[t]
\centering

\caption{Fitting residuals of the reference stellar loci for giants and
dwarfs in different colors.}
\label{tab:locus_residuals}

\setlength{\tabcolsep}{4pt}
\small

\begin{tabular}{lcc}
\hline
Color & Giant & Dwarf \\
\hline

$BP-RP$ & 0.011 & 0.011 \\
$g-r$                   & 0.013 & 0.013 \\
$r-i$                   & 0.013 & 0.013 \\
$i-J$                   & 0.024 & 0.027 \\
$J-H$                   & 0.032 & 0.036 \\
$H-K_s$           & 0.031 & 0.039 \\
$K_s-W_1$          & 0.029 & 0.038 \\

\hline
\end{tabular}

\end{table}

\begin{table*}
\centering
\caption{Polynomial coefficients of the fitted stellar loci for dwarfs and giants.}
\label{tab:locus}
\begin{tabular}{lrrrrrr}
\hline
Color & $A_0$ & $A_1$ & $A_2$ & $A_3$ & $A_4$ & $A_5$ \\
\hline

\multicolumn{7}{l}{Dwarfs} \\

$BP-RP$
& 0.4749 & 1.0096 & 0.0498 & -0.3229 & 0.1952 & -0.0333 \\

$g-r$
& 0.0508 & 0.4560 & 0.0830 & 0.1115 & -0.1562 & 0.0351 \\

$r-i$
& -0.0508 & 0.5440 & -0.0830 & -0.1115 & 0.1562 & -0.0351 \\

$i-J$
& 0.6895 & 0.5158 & 0.2419 & -0.4821 & 0.2296 & -0.0340 \\

$J-H$
& 0.1536 & 0.4801 & 0.1710 & -0.3123 & 0.1041 & -0.0104 \\

$H-K_s$
& 0.0324 & 0.0703 & 0.0424 & -0.0482 & 0.0242 & -0.0042 \\

$K_s-W_1$
& 0.0310 & 0.0215 & -0.0172 & 0.0260 & -0.0062 & 0.0001 \\

\hline

\multicolumn{7}{l}{Giants} \\

$BP-RP$
& 0.4221 & 1.0760 & 0.0684 & -0.2261 & 0.0751 &  \\

$g-r$
& 0.0285 & 0.4664 & 0.1199 & -0.0722 & -0.0069 &  \\

$r-i$
& -0.0285 & 0.5336 & -0.1199 & 0.0722 & 0.0069 &  \\

$i-J$
& 0.6944 & 0.4578 & 0.3731 & -0.3512 & 0.0795 &  \\

$J-H$
& 0.1668 & 0.4968 & 0.0825 & -0.1042 & 0.0133 &  \\

$H-K_s$
& 0.0258 & 0.0471 & 0.1222 & -0.0672 & 0.0103 &  \\

$K_s-W_1$
& 0.0272 & 0.0207 & 0.0587 & -0.0247 & 0.0013 &  \\

\hline
\end{tabular}
\end{table*}

\subsection{The catalog}
\label{sec:the catalog}

Using the SED-fitting procedure described above, we derive individual stellar extinction estimates for approximately 87 million stars. The final catalog contains the Gaia DR3 source identifier, equatorial coordinates (RA and DEC), Galactic coordinates ($l$ and $b$), stellar distances and their uncertainties, and the derived $E(B-V)$ and $A_V$ values with their corresponding uncertainties. The stellar distances are taken from \citet{zhang2025three}. For the reported $A_V$ values, we adopt $R_V\equiv A_V/E(B-V)=3.1$ \citep{fitzpatrick1999correcting}. We also include the minimum reduced chi-square value, $\chi_{\nu}^{2}$, obtained from Equation~\ref{eq:chi2}, as a diagnostic of the agreement between the observed colors and the best-fitting reference SED. A detailed description of the catalog columns is given in Table~\ref{tab:extinction_catalog_description}.

The complete catalog is publicly available at \url{https://nadc.china-vo.org/res/r101905/}.
To facilitate access to this large data set, the catalog is organized into four FITS tables corresponding to the four Galactic quadrants: $0^\circ \leq l < 90^\circ$ (Quadrant I), $90^\circ \leq l < 180^\circ$ (Quadrant II), $180^\circ \leq l < 270^\circ$ (Quadrant III), and $270^\circ \leq l < 360^\circ$ (Quadrant IV).
All four tables share the same column names, column order, data types, units, and column definitions.
A detailed ReadMe is provided at the same website, describing the division criterion, the Galactic longitude ranges, the coordinate systems, and the contents of each column.
Users can download the relevant quadrant tables independently and read them using standard FITS-compatible software, without downloading the entire catalog or running the Python program. Regions crossing quadrant boundaries require the corresponding tables.

At the same website, we provide a standalone Python program as an optional tool for reconstructing the cumulative reddening profile along a user-specified line of sight. Given a Galactic coordinate $(l,b)$, a target distance, and an angular search radius, the program selects surrounding stars from the catalog and fits their distances and $E(B-V)$ measurements to obtain the reddening at the target distance.

\begin{table*}[t]
\centering
\caption{Description of the interstellar extinction catalog. The column definitions and units listed here apply to all four Galactic-quadrant tables.}
\label{tab:extinction_catalog_description}

\setlength{\tabcolsep}{8pt}
\small

\begin{tabular*}{\textwidth}{@{\extracolsep{\fill}}cclc@{}}
\hline
{Column} & {Label} & {Description} & {Unit} \\
  \hline\noalign{\smallskip}
1 & source\_id & Unique source identifier for Gaia DR3 \\
2 & RA & Right Ascension from Gaia DR3 & degrees \\
3 & DEC & Declination from Gaia DR3 & degrees \\
4 & $l$ & Galactic longitude & degrees \\
5 & $b$ & Galactic latitude & degrees \\
6 & d & Distance from \citet{zhang2025three} & kpc \\
7 & d\_err & Error of distance from \citet{zhang2025three} & kpc \\
8 & EBV & E(B-V) from this work & mag \\
9 & EBV\_err & Error of E(B-V) from this work & mag \\
10 & Av & $A_V$ from this work & mag \\
11 & Av\_err & Error of $A_V$ from this work & mag \\
12 & chi2 & Minimum reduced $\chi_{\nu}^{2}$ of the SED fit & -- \\
  \noalign{\smallskip}\hline
\hline
\end{tabular*}

\end{table*}

Following the method of \citet{zhang2026gsed}, we model the cumulative reddening along a given line of sight as a monotonically increasing step function. The selected stars are divided into $K$ distance bins, and the cumulative reddening at a distance $d$ located in the $k$th bin is expressed as

\begin{equation}
E(d)
=
\sum_{j=1}^{k}\exp(h_j),
\qquad
d \in {\rm bin}_k ,
\label{eq:los_extinction}
\end{equation}

where $h_j$ is a free parameter describing the reddening increment in the $j$th distance bin. The exponential form ensures that each incremental reddening, $\Delta E_j=\exp(h_j)$, is positive, thereby guaranteeing that the cumulative reddening increases monotonically with distance.

For a set of $N$ stars selected around a given line of sight, the $i$th star has a measured reddening $E_i$, an uncertainty $\sigma_{E_i}$, and a distance $d_i$. Following \citet{zhang2026gsed}, the model parameters are constrained using a robust Cauchy-like log-likelihood function,

\begin{equation}
\ln \mathcal{L}
=
-\sum_{i=1}^{N}
\ln
\left[
1+
\left(
\frac{
E_i-E_{\rm mod}(d_i,\boldsymbol{\theta})
}{
\sigma_{E_i}
}
\right)^2
\right],
\label{eq:los_likelihood}
\end{equation}

where $E_{\rm mod}(d_i,\boldsymbol{\theta})$ is the cumulative reddening predicted by Equation~\ref{eq:los_extinction} at the distance of the $i$th star, and $\boldsymbol{\theta}$ denotes the set of free parameters $\{h_1,h_2,\ldots,h_K\}$. Compared with a conventional Gaussian likelihood, this logarithmic form reduces the influence of stars with large residuals and therefore makes the inferred extinction profile less sensitive to outliers.

The fitting procedure implemented in the accompanying Python program consists of two stages. First, a non-linear least-squares optimization is performed to locate an appropriate region of parameter space and provide initial estimates of the model parameters. These estimates are subsequently used to initialize a Markov Chain Monte Carlo (MCMC) analysis to sample the posterior distributions of the parameters. The resulting posterior samples are used to determine the best-fitting cumulative reddening profile and its uncertainty. For a target located at a specified distance $d_{\rm target}$, the corresponding reddening can then be obtained directly from the fitted profile as $E(d_{\rm target})$.

\begin{figure*}[t]
\centering
\includegraphics[width=0.95\textwidth]
{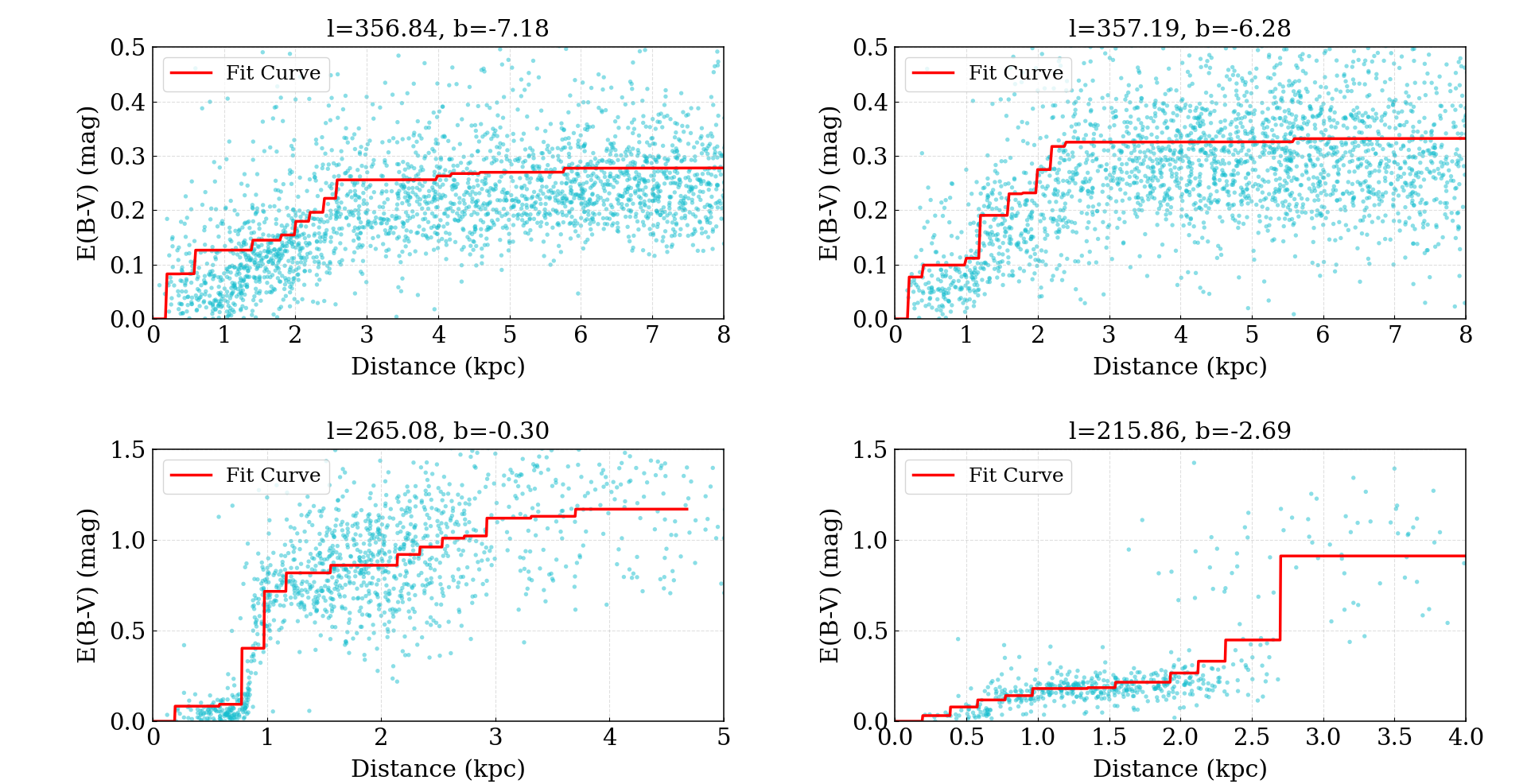}

\caption{Examples of cumulative line-of-sight reddening profiles derived using the Python fitting program provided with the catalog. The individual points represent stars with measured $E(B-V)$ values and distances, while the solid red curves show the best-fitting cumulative reddening profiles. The variations of reddening with distance trace the distribution of interstellar dust along the corresponding lines of sight.}

\label{fig:los_extinction}
\end{figure*}

Figure~\ref{fig:los_extinction} presents representative examples of the line-of-sight extinction profiles obtained with the Python program. The individual stellar measurements trace the variation of $E(B-V)$ with distance, while the solid red curves show the best-fitting cumulative reddening profiles derived using Equations~\ref{eq:los_extinction} and \ref{eq:los_likelihood}. The two upper panels correspond to the two HEALPix pixels containing the largest numbers of stars in our catalog and illustrate the performance of the fitting procedure in densely sampled lines of sight. In both cases, the cumulative reddening increases through a series of distinct steps and gradually approaches a nearly constant value at larger distances.

The lower-left panel corresponds to the direction of the Vela Molecular Ridge, a prominent molecular-cloud complex in the Vela and Puppis regions \citep{murphy1991molecular}. Previous studies place the VMR A, C, and D components at approximately $0.7\pm0.2$~kpc, while the more distant VMR B component lies at about 2~kpc \citep{pettersson2008young}. Our fitted profile shows a pronounced increase in $E(B-V)$ beginning at approximately 0.7~kpc, followed by additional increases at larger distances. The onset of this rise is compatible with the approximate distances of the nearby VMR components, while the reddening continues to increase more gradually farther along the line of sight.

The lower-right panel samples the direction of the Maddalena Molecular Cloud (G216$-$2.5), for which previous studies have reported distances of approximately 2.1--2.2~kpc \citep{megeath2009detection}. In our fitted profile, the reddening remains relatively modest at smaller distances but rises sharply beyond approximately 2.5~kpc, reaching $E(B-V)\simeq0.9$~mag. This increase indicates a substantial dust concentration along the sightline, although its location is farther away than the quoted cloud distance and does not by itself establish an association with the Maddalena cloud. Together, these examples illustrate how the fitted profiles describe both gradual reddening trends and localized increases traced by the stellar measurements.

\subsection{3D Extinction map}
\label{sec:3d map}

To investigate the three-dimensional distribution of interstellar dust, we construct cumulative extinction maps at different distances using the stellar reddening measurements in our catalog. We first select stars with reliable distance estimates by requiring a relative distance uncertainty of $\sigma_d/d<0.2$. The sky is then divided into equal-area pixels using the HEALPix scheme \citep{gorski2005healpix} with $N_{\rm side}=128$, corresponding to a pixel area of approximately $0.21~{\rm deg}^{2}$. For each distance limit, stars satisfying the corresponding distance criterion are assigned to HEALPix pixels according to their Galactic coordinates $(l,b)$, and the characteristic reddening in each pixel is represented by the median $E(B-V)$ value of the selected stars.

\begin{figure*}[t]
\centering
\includegraphics[width=0.49\textwidth]{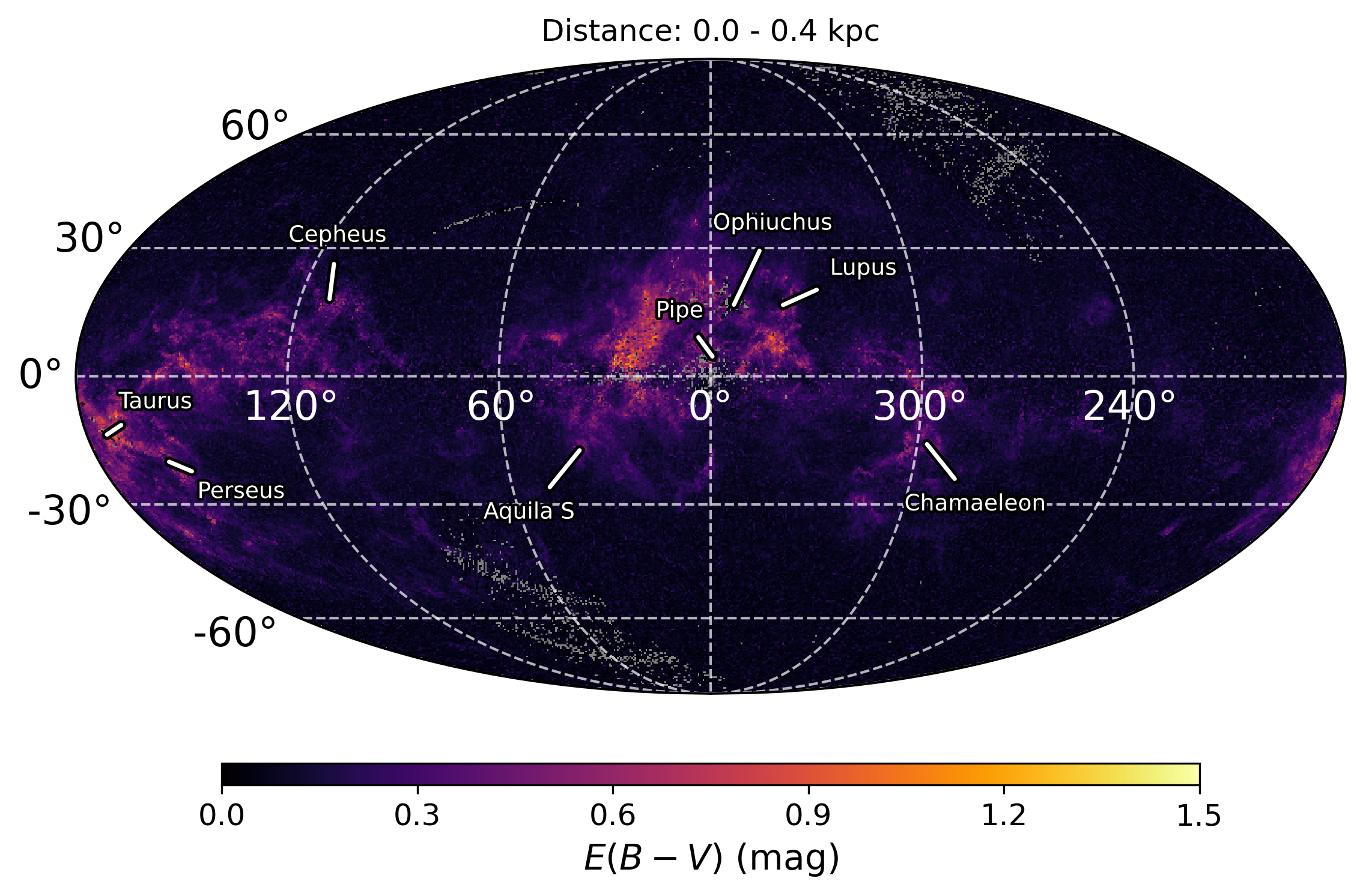}
\includegraphics[width=0.49\textwidth]{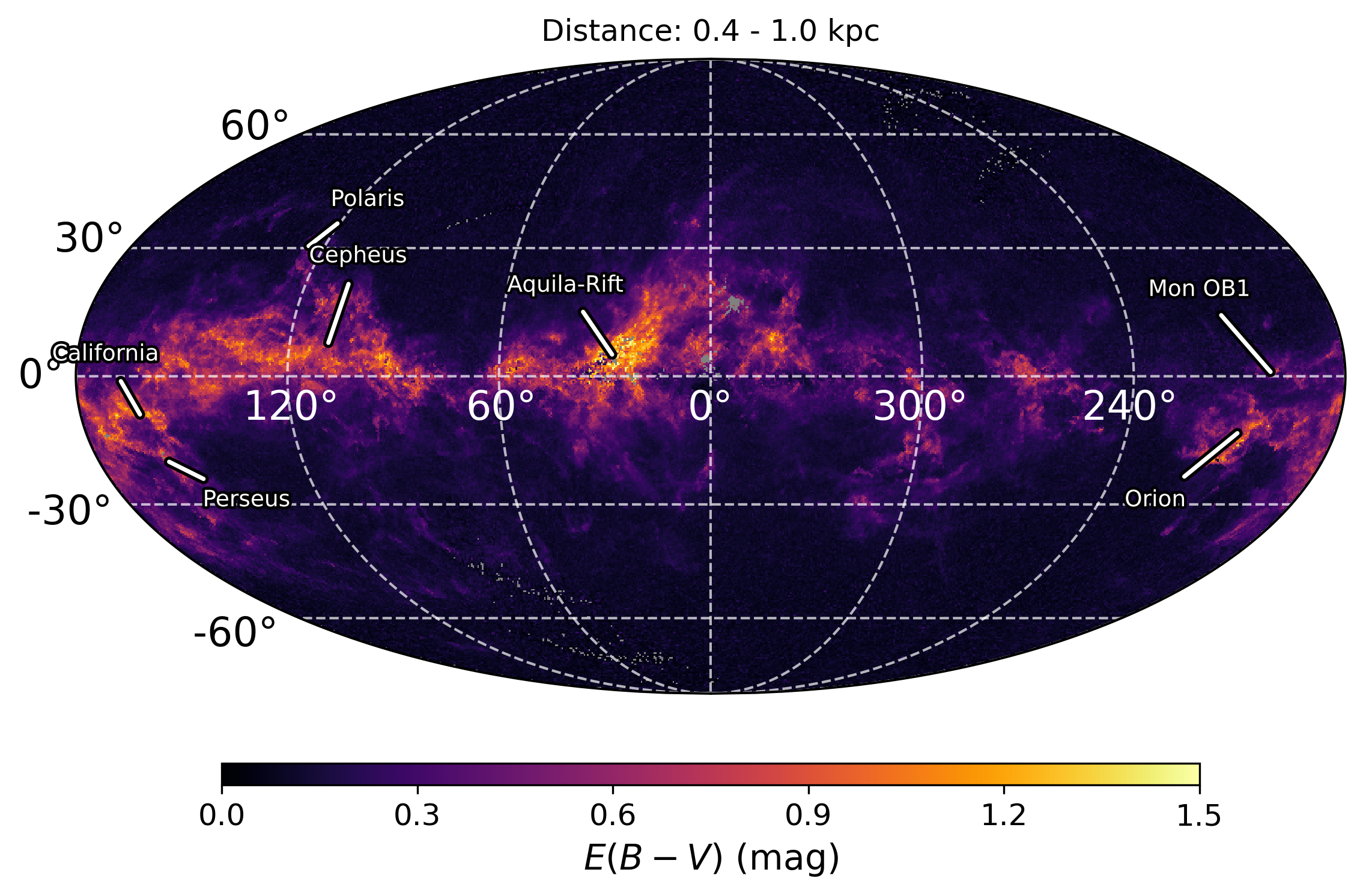}
\includegraphics[width=0.49\textwidth]{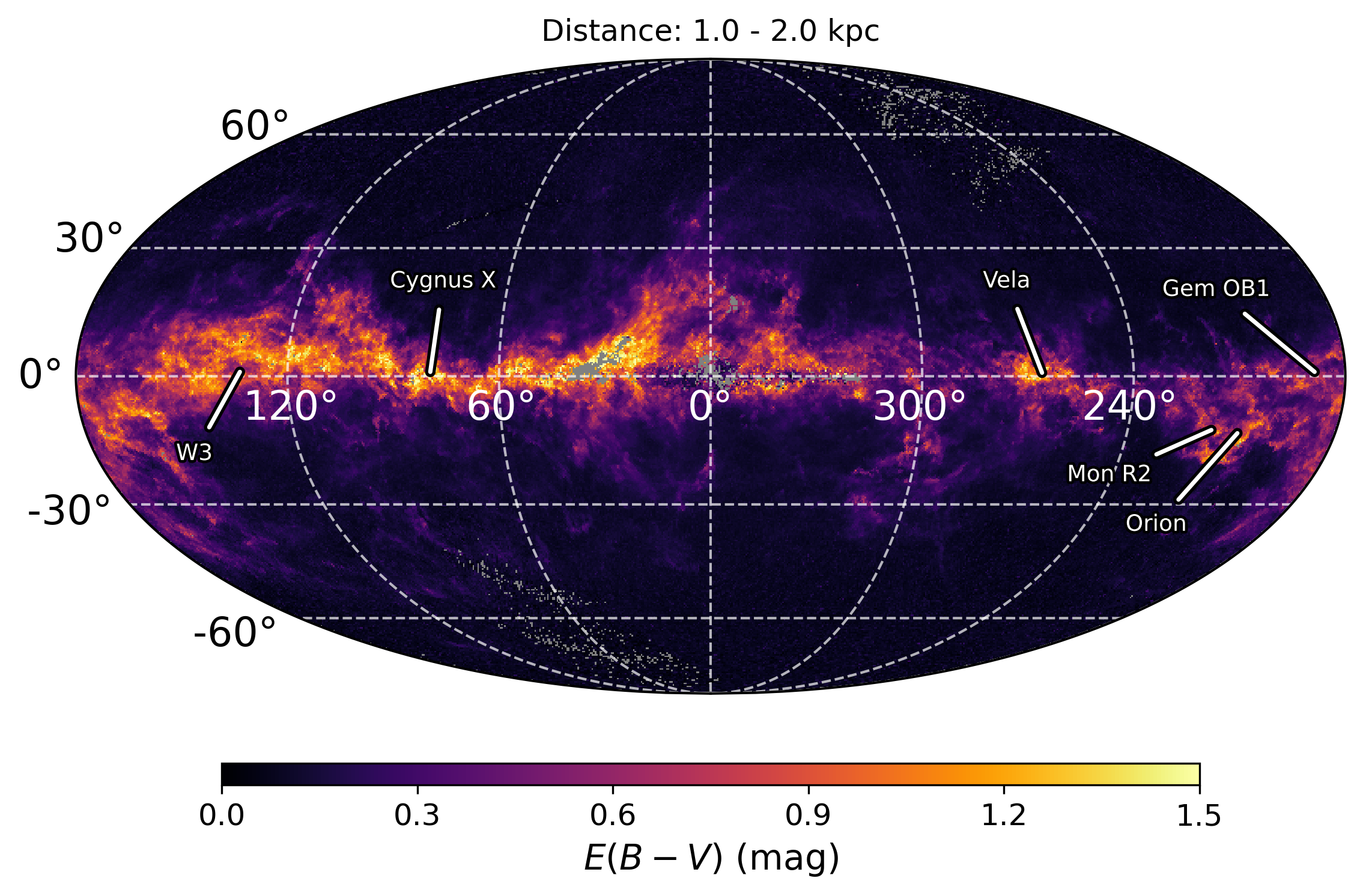}
\includegraphics[width=0.49\textwidth]{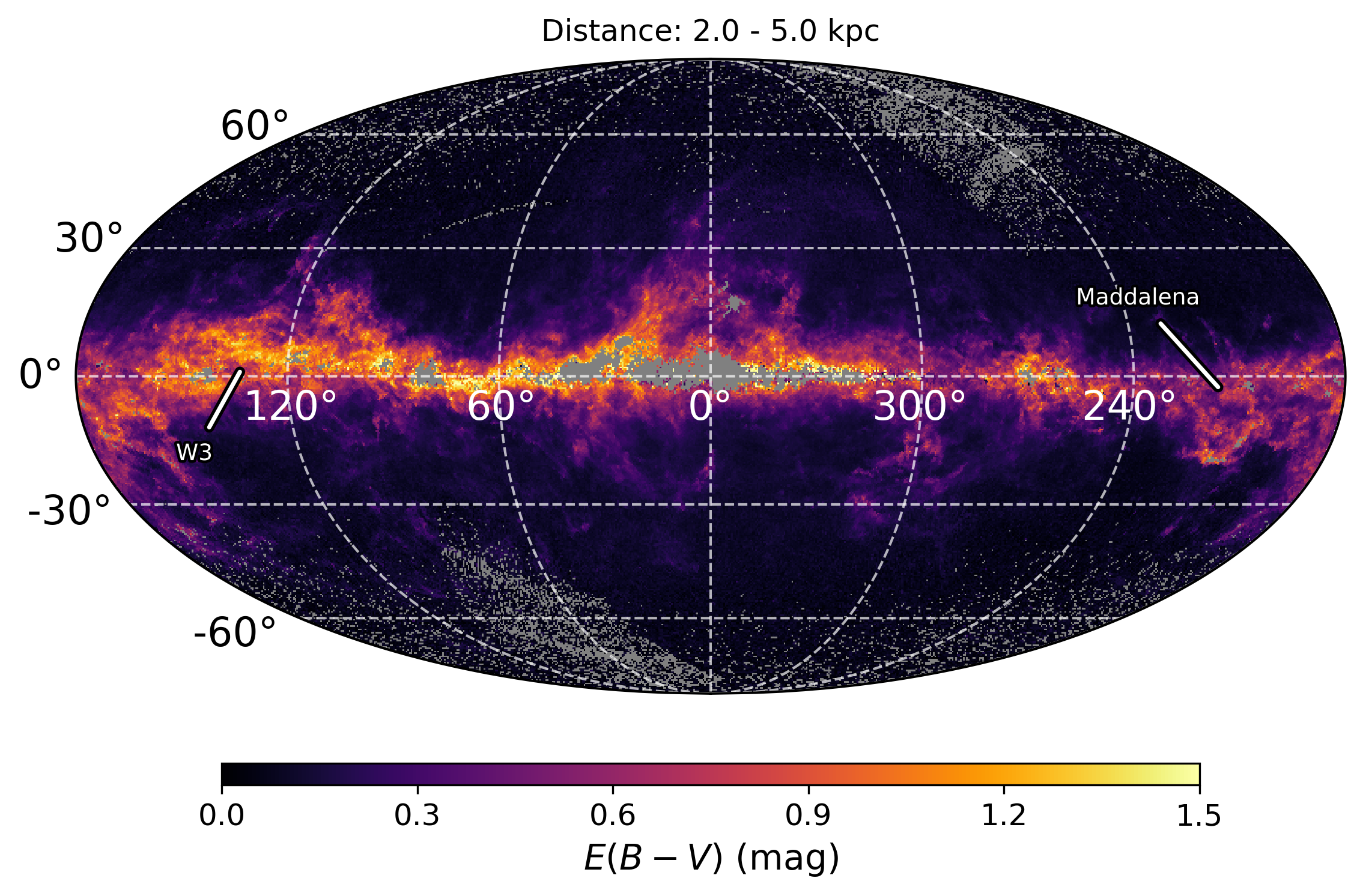}
\caption{Cumulative and bin-sliced $E(B-V)$ reddening sky maps shown in Galactic coordinates with a Mollweide projection ($l = 0^\circ$ centered). The panels display distributions for four discrete distance slices ($0 < d < 0.4~{\rm kpc}$, $0.4 \le d < 1.0~{\rm kpc}$, $1.0 \le d < 2.0~{\rm kpc}$, and $2.0 \le d < 5.0~{\rm kpc}$). The corresponding distance range is marked atop each panel. The color bar denotes $E(B-V)$ in mag, and grey shaded areas represent regions without sample coverage.}
\label{fig:ebv_3d_maps}
\end{figure*}

The full catalog covers a sky area of $41\,246.88~{\rm deg}^{2}$, corresponding to approximately $99.99\%$ of the full sky. Figure~\ref{fig:ebv_3d_maps} presents the cumulative $E(B-V)$ distributions out to distances of 0.4, 1, 2, and 5\,kpc, respectively, in Galactic coordinates using a Mollweide projection. Nearby, localized dust structures dominate at small distances, while more extended features and the concentration of high extinction toward the Galactic plane become apparent at larger distances. Several well-known molecular clouds and star-forming regions are marked for reference, and many coincide spatially with prominent reddening enhancements. The maps thus provide a coherent view of the Galactic dust distribution traced by stars with $\sigma_d/d<0.2$.

\section{Validation and Application}
\label{sec:validation and application}

\subsection{Comparison with Previous Studies}
\label{sec:comparison}

\begin{figure*}[t]
\centering
\includegraphics[width=\textwidth]{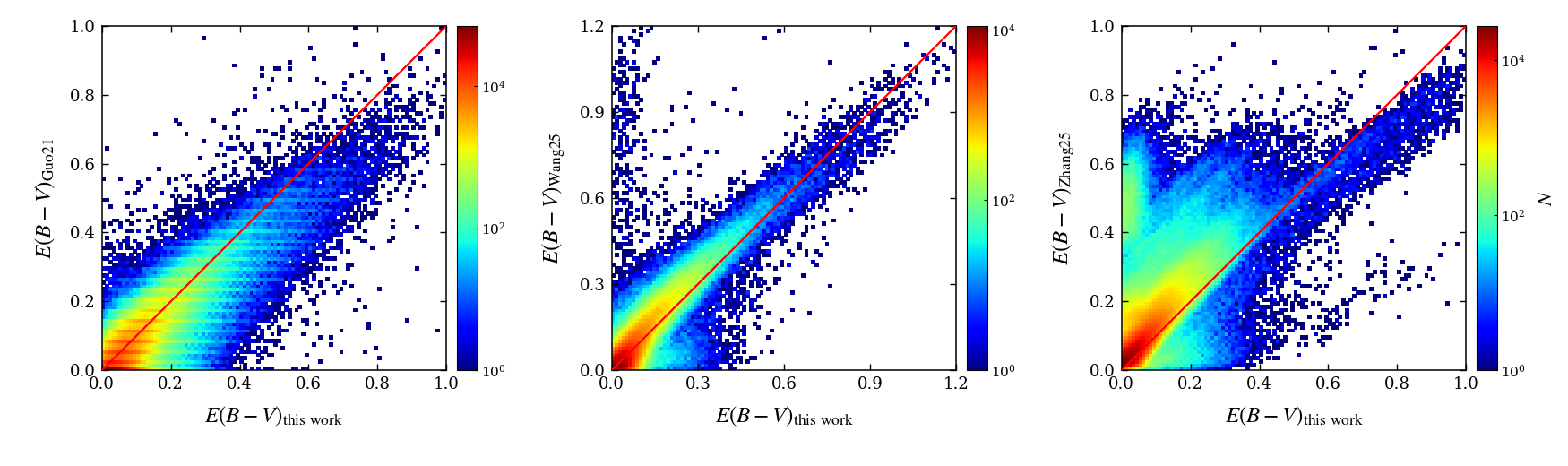}
\caption{Comparison of the $E(B-V)$ values derived in this work with those
from previous studies. From left to right, the three panels show comparisons
with \citet{guo2021three}, \citet{wang2025all}, and \citet{zhang2025three},
respectively. The red solid lines indicate the one-to-one relation, while
the color scale represents the number of stars in each bin.}
\label{fig:ebv_compare}
\end{figure*}

To assess the reliability of our reddening measurements, we compare the derived $E(B-V)$ values with three external stellar extinction catalogs, as shown in Fig.~\ref{fig:ebv_compare}. From left to right, the three panels show the comparisons with \citet{guo2021three}, \citet{wang2025all}, and \citet{zhang2025three}, respectively. Here, the differences are defined as $\Delta E(B-V)=E(B-V)_{\rm literature}-E(B-V)_{\rm this\,work}$. Overall, our measurements show good consistency with the literature values, although the systematic offsets and dispersions differ among the three comparisons.

The left panel compares our results with the reddening estimates of \citet{guo2021three}, who derived stellar extinction using SED fitting based on multi-band photometry from SMSS, Gaia, 2MASS, and WISE. For the common stars, the distribution of the $E(B-V)$ differences has a mean of $\mu=-0.017\,{\rm mag}$ and a standard deviation of $\sigma=0.049\,{\rm mag}$, indicating a small systematic offset and a relatively tight agreement. A noticeable feature in this comparison is the presence of several nearly horizontal bands in the $E(B-V)$ values of \citet{guo2021three}. We attribute these structures to the relatively discrete sampling of $A_r$, or related fitting parameters, adopted during the numerical search for the minimum $\chi^2$. Such discrete sampling can lead to a quantization of the resulting extinction estimates and therefore produce the horizontal banding visible in the left panel. In comparison, the finer parameter sampling adopted in our fitting yields a substantially smoother distribution of reddening estimates.

The middle panel shows the comparison with \citet{wang2025all}, who determined $E(B-V)$ for LAMOST stars using the star-pair method. In this approach, stars with similar spectroscopically determined atmospheric parameters are used to estimate the intrinsic stellar colors, from which the reddening is derived. The comparison yields $\mu=0.010\,{\rm mag}$ and $\sigma=0.049\,{\rm mag}$. The small systematic offset and dispersion indicate that the reddening scale obtained from our photometric SED fitting is highly consistent with that independently determined from the LAMOST spectroscopic sample.

The right panel compares our measurements with those of \citet{zhang2025three}. The corresponding difference distribution has $\mu=0.042\,{\rm mag}$ and $\sigma=0.075\,{\rm mag}$, showing both a larger systematic offset and a larger dispersion than in the other two comparisons. \citet{zhang2025three} adopted a forward-modeling framework in which stellar atmospheric parameters and extinction were inferred simultaneously. In such a joint-fitting procedure, uncertainties in the stellar atmospheric parameters can propagate directly into the extinction estimates because of the degeneracy between intrinsic stellar spectral properties and reddening. In particular, uncertainties in $T_{\rm eff}$, $\log g$, and metallicity can affect the inferred intrinsic spectral energy distribution and consequently the derived $E(B-V)$. This coupling between stellar parameters and extinction likely contributes to the larger scatter and systematic offset seen in the right panel.

\subsection{Application to Extinction-Corrected Color--Absolute-Magnitude Diagrams}
\label{sec:cmd_application}

\begin{figure*}[t]
\centering
\includegraphics[width=0.95\textwidth]{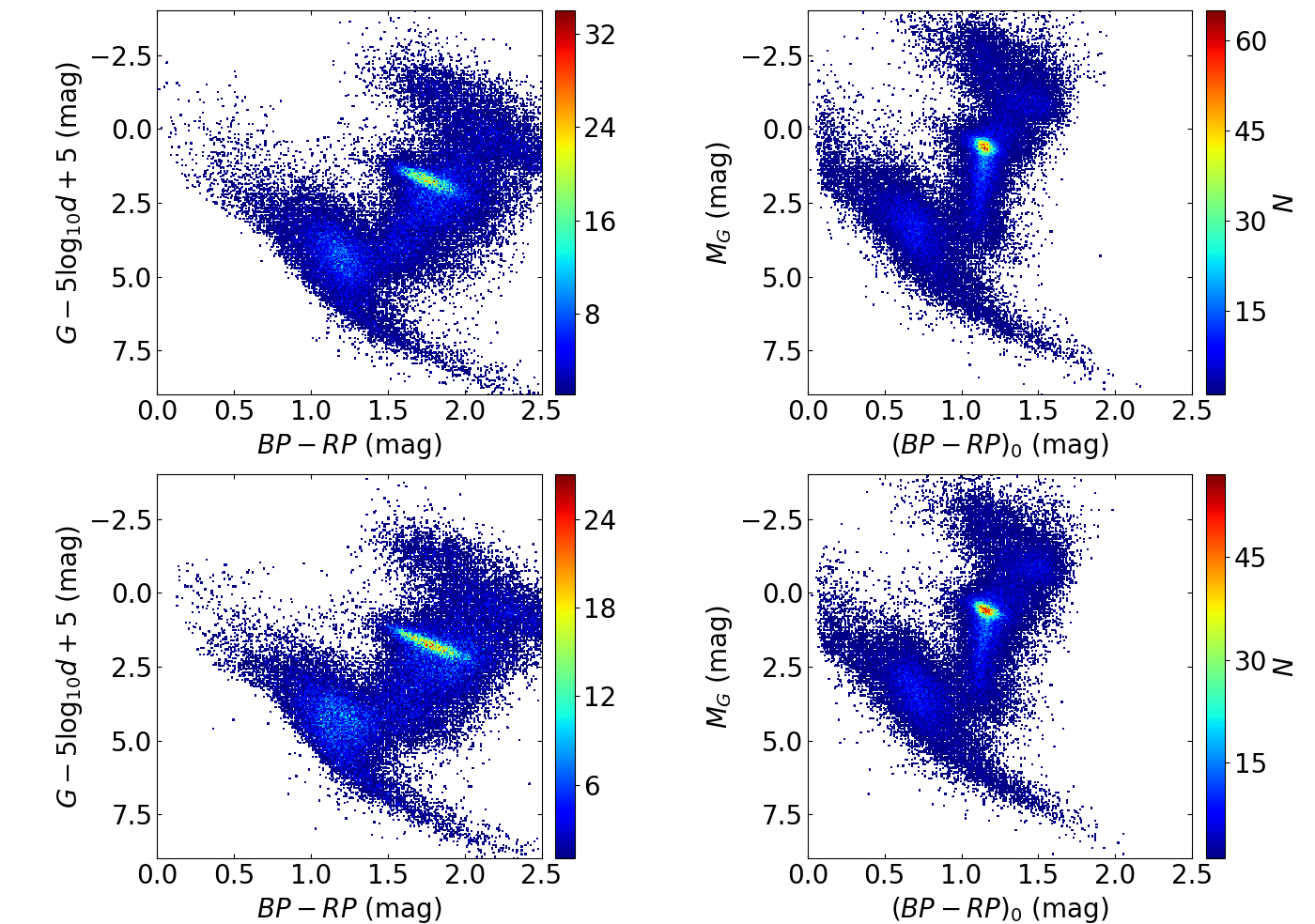}

\caption{Color--absolute-magnitude diagrams of stars in two representative Galactic fields before and after extinction correction. The upper panels correspond to $328^\circ \leq l \leq 330^\circ$ and $-5^\circ \leq b \leq -3^\circ$, while the lower panels correspond to $60^\circ \leq l \leq 62^\circ$ and $2^\circ \leq b \leq 4^\circ$. In each row, the left panel shows the diagram before extinction correction, with $G-5\log_{10}d+5$ plotted against $BP-RP$, while the right panel shows the extinction-corrected diagram, with $M_G$ plotted against $(BP-RP)_0$. The color scale represents the stellar number density. After extinction correction, the stellar sequences become significantly tighter and the red clump feature becomes more compact in both fields.}

\label{fig:cmd}
\end{figure*}

To further demonstrate the reliability and practical usefulness of our stellar extinction estimates, we apply the derived extinction corrections to the color--absolute-magnitude diagrams of stars in two representative Galactic fields, following a similar validation approach adopted in previous extinction studies \citep[e.g.,][]{guo2021three}. The two selected regions are $328^\circ \leq l \leq 330^\circ$, $-5^\circ \leq b \leq -3^\circ$, and $60^\circ \leq l \leq 62^\circ$, $2^\circ \leq b \leq 4^\circ$. The resulting diagrams are shown in Fig.~\ref{fig:cmd}. In each row, the left panel shows the color--absolute-magnitude diagram before extinction correction, while the right panel shows the corresponding diagram after correcting both the color and absolute magnitude for interstellar extinction. The upper and lower rows correspond to the two selected sky regions, respectively.

A comparison between the left and right panels clearly shows that the extinction correction significantly reduces the broadening of the stellar sequences in both fields. After correction, prominent evolutionary features, including the main sequence, red giant branch, and red clump (RC), become more clearly defined. In particular, the RC feature becomes substantially more compact in the extinction-corrected color--absolute-magnitude diagrams, indicating that our extinction estimates effectively reduce the effects of differential reddening and line-of-sight extinction.

To further quantify the effectiveness of the extinction correction, we determine the density peak of the RC feature in the corrected color--absolute-magnitude diagrams. For both selected regions, the RC peak is located at $(BP-RP)_0 = 1.1625\,{\rm mag}$ and $M_G = 0.65\,{\rm mag}$. The nearly identical RC peak positions obtained for the two independent fields indicate that the extinction correction yields a consistent intrinsic RC locus along different lines of sight. The derived RC position is also broadly consistent with expectations from stellar evolution models, such as the PARSEC isochrones \citep{bressan2012parsec}. These results provide an additional validation of the reliability of our extinction estimates and demonstrate their applicability to studies of intrinsic stellar populations.

\section{Conclusions}
\label{sec:conclusions}

We present a catalog of $E(B-V)$ estimates and their uncertainties for 87,270,425 stars, based on synthetic SMSS $g$, $r$, and $i$ photometry from corrected Gaia DR3 XP spectra, combined with Gaia BP and RP, 2MASS $JHK_s$, and WISE $W_1$ photometry. Extinction is determined by minimizing $\chi_{\nu}^{2}$ against empirical stellar loci constructed separately for dwarfs and giants using 1,755,077 low-extinction reference stars. The catalog covers $41\,246.88~{\rm deg}^{2}$, or approximately $99.99\%$ of the full sky, and provides a homogeneous data set for stellar extinction correction and studies of Galactic dust.

A Python program accompanying the catalog reconstructs cumulative line-of-sight reddening profiles using a monotonically increasing model and a robust likelihood. We also construct cumulative and bin-sliced reddening maps in four distance intervals spanning 0--5~kpc using stars with $\sigma_d/d<0.2$. These maps reveal prominent molecular clouds and the large-scale concentration of interstellar dust toward the Galactic plane.

Comparisons with three external stellar extinction catalogs support the reliability of our measurements. The differences relative to \citet{guo2021three}, \citet{wang2025all}, and \citet{zhang2025three} have mean values of $-0.017$, $0.010$, and $0.042\,{\rm mag}$ and dispersions of $0.049$, $0.049$, and $0.075\,{\rm mag}$, respectively. The agreement with the independent LAMOST star-pair measurements further supports our reddening scale. Applying the extinction corrections to two Galactic fields produces tighter stellar sequences and a consistent red-clump peak at $(BP-RP)_0=1.1625\,{\rm mag}$ and $M_G=0.65\,{\rm mag}$ in both fields.

Extending this method to larger, better-calibrated samples of Gaia XP spectra could improve the spatial sampling of stellar extinction. Combined with improved stellar distances and complementary multi-band photometry, such extensions could support more detailed studies of molecular clouds and the three-dimensional distribution of Galactic dust. Improvements in the precision of individual extinction estimates will also depend on the quality of the spectral and photometric calibrations and the available multi-band measurements.

\section*{Acknowledgement}
This work is partially supported by the Physics Discipline Construction Project of Chuxiong Normal College (Grant No. 20260600016) and the Chuxiong Normal College Doctoral Research Initiation Fund Project under No. BSQD2586.

This work has made use of data from the European Space Agency (ESA) mission Gaia (\url{https://www.cosmos.esa.int/gaia}), processed by the Gaia Data Processing and Analysis Consortium (DPAC, \url{https://www.cosmos.esa.int/web/gaia/dpac/consortium}). Funding for the DPAC has been provided by national institutions, in particular the institutions participating in the Gaia Multilateral Agreement.

This publication makes use of data products from the Two Micron All Sky Survey, which is a joint project of the University of Massachusetts and the Infrared Processing and Analysis Center/California Institute of Technology, funded by the National Aeronautics and
Space Administration and the National Science Foundation.

This publication makes use of data products from the Widefield Infrared Survey Explorer, which is a joint project of the University of California, Los Angeles, and the Jet Propulsion Laboratory/California Institute of Technology, funded by the National Aeronautics and Space Administration.

\bibliographystyle{cas-model2-names}
\bibliography{bibtex}

\end{document}